\documentclass[aps,pra,twocolumn,superscriptaddress,showpacs,floatfix]{revtex4}
\usepackage{graphicx,amsmath,amssymb,xspace,epsfig,float}

\begin{document}

\title{Fermionization of a strongly interacting Bose-Fermi mixture in a one-dimensional harmonic trap}

\author{Bess Fang}
\affiliation{Department of Physics, Block S12, Faculty of Science, National
  University of Singapore, 2 Science Drive 3, Singapore 117542}
\affiliation{Centre for Quantum Technologies, National University of
  Singapore, 3 Science Drive 2, Singapore 117543}

\author{Patrizia Vignolo}
\affiliation{Institut Non Lin\'eaire de Nice, Universit\'e de Nice-Sophia
  Antipolis, CNRS, 1361 route des Lucioles, 06560 Valbonne, France}

\author{Christian Miniatura}
\affiliation{Department of Physics, Block S12, Faculty of Science, National
  University of Singapore, 2 Science Drive 3, Singapore 117542}
\affiliation{Centre for Quantum Technologies, National University of
  Singapore, 3 Science Drive 2, Singapore 117543}
\affiliation{Institut Non Lin\'eaire de Nice, Universit\'e de Nice-Sophia
  Antipolis, CNRS, 1361 route des Lucioles, 06560 Valbonne, France}
\affiliation{IPAL, CNRS, 12R, 1 Fusionopolis Way, Singapore 138632, Singapore}
\author{Anna Minguzzi}
\affiliation{Universit\'e Joseph Fourier, Laboratoire de Physique et
  Mod\'elisation, des Milieux Condens\'es, CNRS B.P.\ 166, 38042 Grenoble,
  France}

\date{\today}

\begin{abstract}
  We consider 
 a strongly interacting one-dimensional (1D) Bose-Fermi mixture confined in a harmonic trap. It consists of a Tonks-Girardeau (TG) gas (1D Bose gas with
  repulsive hard-core interactions) and of a non-interacting Fermi gas (1D spin-aligned Fermi
  gas), both species interacting through hard-core repulsive interactions. Using a generalized Bose-Fermi mapping, we determine the one-body density matrices, exact particle density profiles, momentum distributions and behaviour of the mixture under 1D expansion when opening the trap. In real space, bosons and fermions do not display any phase separation: the respective density profiles extend over the same region and they both 
  present a number of peaks equal to the total number of particles in
  the trap.   In momentum space the bosonic component has the typical
  narrow TG profile, while the fermionic component shows a broad
  distribution with fermionic
  oscillations at small momenta. Due to the large boson-fermion
  repulsive interactions, both the bosonic and the fermionic
  momentum distributions decay as   
  $C p^{-4}$
  at large momenta,  like in the case of a pure bosonic TG gas. The coefficient $C$ is related to the two-body density matrix 
and to the bosonic concentration in the mixture. 
When opening the trap, both momentum distributions "fermionize" under expansion and turn into that of a
  Fermi gas with a particle number equal to the total number of particles in
  the mixture. 
\end{abstract}


\maketitle

\section{Introduction}
\begin{figure}
 \includegraphics[width=0.65\linewidth,clip=true]{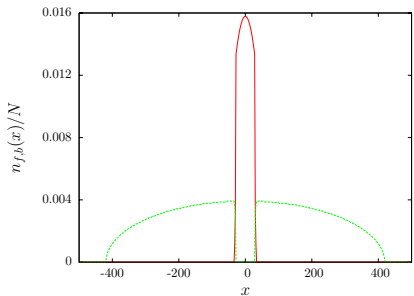}
 \caption{\label{fig_intro} (Color online) Bosonic (red solid line) and 
fermionic ($\times 20$, green dashed line)
density profiles (in units of $1/\ell_0$ and normalized to the total number $N$ of particles) as a function of the distance $x$ from the trap center (in units of
$\ell_0$) as predicted by a 
two-fluid mean-field model \cite{Akdeniz2005} for a BF mixture confined in a 1D harmonic trap.
The length scale unit $\ell_0$ is the harmonic oscillator length as defined in Sec. \ref{sec:model}.
%$N$ is the total number of particles.
The boson-boson and boson-fermion interaction strengths have been choosen such that 
$\gamma_{BB}=1.5\times 10^{-4}$ and 
$\gamma_{BF}=0.63$ (see the text for the definition of $\gamma_{BB}$ and 
$\gamma_{BF}$).
}
\end{figure}

Dilute degenerate Bose-Fermi (BF) or Fermi-Fermi mixtures have been realized over the past few years in several experiments
by trapping and cooling either gases made of mixed alkali-atom isotopes 
\cite{Schreck2001,Goldwin2002,Hadzibabic2002,Modugno2002,Ospelkaus2006,Dieckmann2008},
or of imbalanced two-spin fermionic atoms \cite{Partridge2006,Zwierlein2006}. In the latter, a superfluid paired core 
is surrounded by a shell of normal unpaired fermions. For BF mixtures a mean-field approach predicts that the 
boson-fermion coupling can lead to 
quantum phase transitions and, 
in particular, that boson-fermion repulsion can induce a spatial demixing of 
the bosonic and fermionic components
when the interaction energy overcomes the kinetic and confinement 
energies \cite{Viverit2000}.
The instability leading to this phase separation has been studied 
in all dimensions both for homogeneous 
\cite{Viverit2000,Das2003} and for harmonically-trapped mixtures 
\cite{Akdeniz2002,Akdeniz2004,Akdeniz2005}.
As an illustrative example, we show in Fig. \ref{fig_intro} the
bosonic and fermionic density profiles $n_B(x)$ and $n_F(x)$ as obtained from a two-fluid mean-field model when a BF mixture with equal masses $m$ is confined in a 1D harmonic trap and phase separation occurs as the Pauli pressure overcomes the boson-boson repulsion \cite{Akdeniz2005}. In the model, the particles experience contact boson-boson  $v_{BB}(x) = g_{BB} \delta(x)$ and boson-fermion  $v_{BF}(x) = g_{BF} \delta(x)$ interactions. The quasi-1D 
interaction  strengths $g_{BB}$ and $g_{BF}$, which can be expressed in terms of the 3D scattering lengths  \cite{Olshanii1998}, define in turn the  adimensional coupling constants $\gamma_{BB}=\hbar^2 g_{BB}/(m n_{B})$ and
$\gamma_{BF}=\hbar^2 g_{BF}/(m n_{F})$. These parameters
measure the ratio of the
interaction to kinetic energies.
Rather counter-intuitively, in 1D, the weakly interacting regime corresponds to the large-density  regime. 
The mean-field description applies well to {\em local} observables in the weakly interacting regime,
however, because phase fluctuations do play a major role in 1D, the off-diagonal long-range order is lost 
and the one-body density matrix
decays as a power-law, even at zero temperature, with an exponent governed by the strength of the interactions  (see e.g. \cite{Giamarchi2003}).

In this work, we focus on the strongly interacting regime
$\gamma_{BB}\gg 1$, $\gamma_{BF}\gg 1$ which is beyond the regime of validity of the
mean-field treatment.
This strongly interacting regime is not far from experimental access. Indeed the increasing sophistication of experimental techniques allows to realize
traps so tight that the atomic dynamics is essentially one-dimensional
\cite{Moritz2003} and to drive the effective 1D 
coupling strength to very large values 
%(or equivalently $a$ to very low values) 
\cite{Paredes2004,Kinoshita2004}. 
%can allow in principle the experimental observation or 
%the experimental lack of phase separation of a BF mixture in
%this geometry.
Previous theoretical studies of strongly interacting BF mixtures include the   
extension of
the Bethe Ansatz method developed by Lieb and Lininger for bosons to the homogeneous 
BF mixture (see e.g.,\cite{Imambekov2006} and references therein), methods from conformal quantum field theory \cite{Frahm2005}, a numerical analysis of
a harmonically trapped Bose-Bose mixture
using a multi-configuration time-dependent Hartree method
\cite{Zoellner2008} 
and the exact solution for the many-body wavefunction using a generalized
Bose-Fermi mapping method valid when the Tonks-Girardeau (TG) limit $\gamma_{BB} \to \infty$ and $\gamma_{BF} \to \infty$ is obtained for both species 
\cite{Girardeau2007}.

In this paper we analyze several equilibrium and dynamical properties of the inhomogeneous Bose-Fermi mixture in the TG regime. We use the TG many-body wavefunction determined  in Ref.~\cite{Girardeau2007}  to obtain
the exact density profiles, momentum distributions and behaviour under longitudinal expansion  of a mixture of bosons and fermions (with equal masses $m$) trapped by the same external harmonic potential. The bosonic component of the mixture is made of identical impenetrable bosons (TG gas) and the fermionic component is made of non-interacting identical spin-polarized fermions. Finally the boson-fermion interaction is characterized by a point-like infinite hard-core repulsion. As a whole the BF mixture under consideration is thus made of impenetrable particles. In the TG limit, at variance with the mean-field predictions,   
 this system does not  
display any phase separation in real space: the boson and fermion density 
profiles are proportional to each other, they extend over the same
region of space and both  
present a number of density peaks equal to the total number of particles 
in the trap.
The bosons and fermions display however differences in their
momentum distributions, especially at small momenta where the bosonic
momentum distribution is sharply peaked around $p=0$ while the fermionic
one, reflecting the Fermi sphere occupancy, is broad. Quite
remarkably, we also find that {\em both}
momentum distributions  
have the same $p^{-4}$ asymptotic behaviour as the pure hard-core Bose gas \cite{Minguzzi2002,Olshanii2003}. We relate the coefficient of this power-law decay to the 2-body
density matrix of the system.
For the fermionic component, this "bosonization'' of the momentum 
distribution is due to the hard-core boson-fermion repulsion.
We also study the 1D expansion of the mixture when the trapping external potential is turned off and the BF mixture is released from the trap and is allowed to expand in a 1D waveguide. The net result of this expansion is to "fermionize" the initial fermion and boson momentum distributions: 
once the expansion is completed, the shape of the resulting distributions in momentum space is the same as the shape of the initial ones in real space, and  they both present  
a number of peaks equal to the total number of particles. Our work thus generalizes the results obtained for a pure TG gas in \cite{Rigol2005,Minguzzi2005}. 

The paper is organized as follows. Section \ref{sec:model} introduces
the model for the BF mixture under consideration.
In Sec. \ref{sec:onebody} we derive the bosonic 
and fermionic
one-body density matrices and, in 
Secs. \ref{sec:density} and \ref{sec:momentum} respectively, we compute analytically the corresponding spatial densities and momentum distributions. The effects of the expansion dynamics on the momentum distributions when the trap is turned off are described in Sec. \ref{sec:exp}. Finally Section \ref{sec:concl} offers some concluding remarks and perspectives.

\section{\label{sec:model}The model}

We consider a mixture of $N_B$ identical bosons with TG interactions and $N_F$ identical non-interacting fermions having equal masses $m$ and trapped in the same 1D harmonic potential with trapping frequency $\omega_0$. Every boson-fermion pair is subjected to TG (repulsive) interactions. Throughout the paper we will use dimensionless variables, expressing all spatial variables in units of the harmonic oscillator length $\ell_0 = \sqrt{\hbar/m\omega_0}$, all momenta in units of the harmonic oscillator momentum $p_0 = \hbar/\ell_0 = \sqrt{m\hbar\omega_0}$, all energies in units of the harmonic oscillator energy $\hbar\omega_0$ and time in units of $\omega_0^{-1}$. By convention, we will collectively denote the space coordinates by $X = (x_1, ..., x_N)$ understanding that $x_i$ is a scaled bosonic variable provided $i \in \mathcal{B} = \{1, ..., N_B\}$ while it is a scaled fermionic one provided $i \in \mathcal{F} = \{N_B+1, ..., N_B+N_F=N\}$. The same convention will apply for
  scaled momenta denoted collectively as $P = (p_1, ..., p_N)$. 

In the Tonks-Girardeau limit, when $\gamma_{BB}\to \infty$ and
$\gamma_{BF}\to \infty$, the interactions are appropriately incorporated in the model by imposing boundary constraints to the total wavefunction.
In this case of infinite repulsive contact interactions, and in terms of the scaled variables, the total Hamilton differential operator is then a sum of one-body Hamilton differential operators $\hat{H} = \sum_{i=1}^N \hat{h}(i)$ where $\hat{h}(i) = (-\partial_{x_i}^2 +x_i^2)/2$ in space representation.
We now demand that $\Psi(X) = 0$ whenever $x_i = x_j$ for $i\not=j$.
Using the Fermi-Bose mapping theorem \cite{Girardeau2007}, the total wavefunction satisfying these constraints is constructed from a Slater determinant and reads 
\begin{equation}
\Psi (X) = A (X)\, \Psi_D (X).
\end{equation}
Here the antisymmetrizor is given by  
\begin{equation}
\label{eq:A}
 A(X) =\Big[\!\!\!\!\prod_{\begin{array} {c} \scriptstyle
 j,\ell \in \mathcal{B}\\[-1ex] \scriptstyle j<\ell\end{array}} 
\mathrm{sgn}(x_j-x_\ell)\Big]\Big[\prod_{j \in \mathcal{B}}\prod_{\ell \in \mathcal{F}} 
\mathrm{sgn}(x_j-x_\ell)\Big]
\end{equation}
and the determinantal wave function is
\begin{equation}
\Psi_D (X)  =  \sum_P \varepsilon(P)\, \prod_{i =1}^N \, u_i(Px_i), 
\end{equation}
where $P$ runs over all $N!$ 
possible permutations, including boson-fermion exchanges,  $\varepsilon(P)=\pm 1$ is the sign of the permutation and the $\{u_i\}$ are the single-particle orbitals for the given external potential. We recall that the choice (\ref{eq:A}) for the mapping function $A(X)$ is not unique as in the limit $\gamma_{BB}\to \infty$ and $\gamma_{BF}\to\infty$ the ground state has a large degeneracy. We choose here the solution which has the same cusps as the lowest-energy solution for finite interaction strength \cite{Girardeau2007}.

For the harmonic potential the groundstate wave function is thus obtained by taking the rescaled
Hermite-Gaussian orbitals
$\phi_n(x)= (\sqrt{\pi} \, 2^nn!)^{-1/2} \, e^{-x^2/2} \, H_n(x)$. 
As already shown in Ref. \cite{Girardeau2007}, the full many-body groundstate  wavefunction  can then  be written as 
\begin{equation}
\Psi_0(X) = c_N \, e^{-\sum_i x_i^2/2} \, f_0(X)
\end{equation}
 with 
\begin{equation}
f_0 (X) = \Big[\!\!\!\!\prod_{\begin{array} {c} \scriptstyle j,\ell \in 
\mathcal{B}\\[-1ex] \scriptstyle \ell>j\end{array}} |x_j-x_\ell|\Big]
\Big[\!\!\!\!\prod_{\begin{array} {c} \scriptstyle j,\ell 
\in \mathcal{F}
\\[-1ex] \scriptstyle  
\ell>j\end{array}} (x_j-x_\ell)\Big] 
\Big[\prod_{\begin{array} {c} \scriptstyle j \in \mathcal{B}\\[-1ex] \scriptstyle 
\ell \in \mathcal{F}\end{array}}
|x_j-x_\ell|\Big]
\label{thiseq}
\end{equation}
and the $N$-dependent normalization constant \( c_N = 2^{N(N-1)/4} \left[ \pi^N \, N! \prod_{n=0}^{N-1} n!  \right]^{-1/2} \). One can easily check that $\Psi_0$ indeed fulfills the constraints imposed by quantum statistics and the infinite repulsive contact interactions. It is duly \textit{symmetric} under the exchange of any two bosons, it is duly \textit{antisymmetric} under the exchange of any two fermions and it vanishes as soon as any two particles are at the same location in space. Furthermore it is straightforward to check that $|\Psi_0|^2$ is normalized to unity, i.e. $\int dX |\Psi_0(X)|^2 =1$.
  %, \( l_0 = \sqrt{\hbar/m\omega_0} \) being the harmonic oscillator length
%Starting from Eq. (\ref{thiseq}) we denote the rescaled variables as 
%$\tilde x=x/l_0$.

\section{One-body density matrices}
\label{sec:onebody}
The one-body bosonic density matrix $\rho_B^{(1)}$ is obtained by tracing out all particles except one boson. Since $\Psi_0$ is symmetric under the exchange of bosons, it can always be written as
\begin{equation}
\label{rho1def}
\rho_B^{(1)}(x,y)  = N_B  \int dX' \ \Psi_0(x, X') \Psi_0^*(y,X')
\end{equation}
where $X'$ is a shorthand for all other variables $x_i$ except the first bosonic one (i.e. $i = 2, \hdots, N$). Since $\Psi_0$ is normalized to unity, $\rho_B^{(1)}$ is normalized to the total number of bosons $N_B$.
It can be shown that $\rho_B^{(1)}$ is in fact proportional to 
the one-body density matrix of a pure TG gas of $N$ bosons 
\cite{Girardeau2007}, namely
$\rho_B^{(1)}(x,y)=(N_B/N) \rho_{TG} (x,y)$ \cite{Forrester2003} so that we finally get
\begin{equation}
\rho_B^{(1)} (x,y)= \frac{2^{N-1}N_B}{\sqrt{\pi}\, N!} e^{-(x^2+y^2)/2} \, \mathrm{det} \Gamma(x,y)
\end{equation}
where $\Gamma(x,y)$ is a square matrix of order $(N-1)$ with entries
\begin{equation}
\Gamma_{jk}(x,y)=\sqrt{\frac{2^{j+k-2}}{\pi \, \Gamma(j) \,\Gamma(k)}} \! \int_{-\infty}^{\infty}\!\! {\rm d} t\, e^{- t^2} |x- t|\,|y- t| \,t^{j+k-2},
\end{equation}
and $\Gamma(j)=(j-1)!$ is the Gamma function.

The determination of the fermionic one-body density matrix $\rho_F^{(1)}(x,y)$ follows the same rationale, namely integrate over all particles in the mixture except one fermion (which can be chosen to be the last one, i.e. $j=N$). It has been calculated in \cite{Imambekov2006,Girardeau2007} in the case of a homogeneous mixture and the calculation proves more complex. We have extended the result derived in \cite{Girardeau2007} to the harmonically trapped mixture and we find:
\begin{equation}
\label{rhofermion}
\rho_F^{(1)} (x,y)= \frac{2^{N-1} \, N_F! \, N_B!}{\sqrt{\pi}\,N!\,(N-1)!} \, e^{-(x^2+y^2)/2} \, \Delta(x,y)
\end{equation}
where
\begin{equation}
\label{Delta}
\Delta(x,y) = \int_0^{2\pi} \frac{d\phi}{2\pi} e^{-i\phi N_B} \, \mathrm{det} \! \left[\Gamma(x,y)e^{i\phi}+\Lambda(x,y) \right]
\end{equation}
with $\Lambda(x,y)$ being a square matrix of order $(N-1)$ with entries given by the Gaussian integrals
\begin{equation}
\Lambda_{jk}(x,y)=\sqrt{\frac{2^{j+k-2}}{\pi \Gamma(j) \Gamma(k)}} \! \int_{-\infty}^{\infty}\!\!{\rm d} t\, e^{- t^2} (x- t)(y- t) \,t^{j+k-2}.
\end{equation}
The integration over the phase $\phi$ in \eqref{Delta} ensures that only the terms
involving $N_B$ bosons are picked up and the factorials in \eqref{rhofermion} take care
of multiple-counting. Equation \eqref{rhofermion} could not be reduced further to a simpler analytical formula but the preceding expressions allow  for an efficient numerical computation of $\rho_F^{(1)}$.

It is possible to study the large-distance off-diagonal behaviour of the one-body density matrix by factoring out the Gaussian functions imposed by the trap. In practice,  we rescale the bosonic and fermionic one-body density matrices by the square root of their corresponding density profiles, i.e. we define the one-body correlators as in \cite{Schmidt2007}, 
%\begin{eqnarray}
%\label{correlators}
%g_B^{(1)}(x,y)&=\frac{\rho_B^{(1)}(x,y)}{\sqrt{n_B(x)n_B(y)}} = \frac{\mathrm{det} \Gamma(x,y)}{\sqrt{\mathrm{det} \Gamma(x,x)\mathrm{det} \Gamma(y,y)}}, \nonumber \\
 %g_F^{(1)}(x,y)&=\frac{\rho_F^{(1)}(x,y)}{\sqrt{n_F(x)n_F(y)}} = \frac{\Delta(x,y)}{\sqrt{\Delta(x,x)\Delta(y,y)}}.
 % \end{eqnarray} 
 \begin{equation}
\label{correlators}
g^{(1)}(x,y)=\frac{\rho^{(1)}(x,y)}{\sqrt{n(x)n(y)}}.
 \end{equation} 

\begin{figure}
\includegraphics[width=1\linewidth,clip=true]{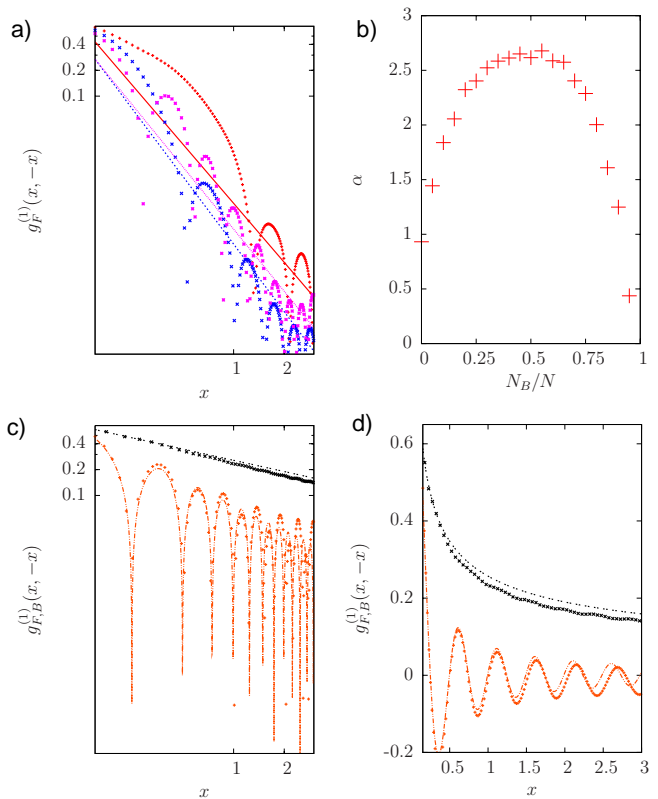}
 \caption{\label{fig0}
(Color online) Plot of the off-diagonal behaviour of the fermionic and bosonic one-body density correlators $g^{(1)}_F(x,-x)$ and $g^{(1)}_B(x,-x)$, Eq. \eqref{correlators}, as a function of the spatial coordinate $x$ (in units of $\ell_0$). 
Panel (a) shows $g^{(1)}_{F}$ for the BF mixture (in log-log scale) with $N=20$ particles and three different bosonic concentrations $N_B/N=.75\;(+), .5\;(\ast), .25\; (\Box)$. The comparison to the fitting function $\sim|x|^{-\alpha}$ (continuous, short-dashed, dotted) allows to extract the slope.  Panel (b) shows the results for the power-law exponent $\alpha$ obtained at large distances $x$ for the mixture with $N=20$ particles when the boson concentration $N_B/N$ is varied. Panel (c) shows  the log-log plots of $g^{(1)}_{B}$ for a pure TG gas with $N=N_B=20$ particles ($\ast$) and of $g^{(1)}_{F}$ for the pure Fermi gas with $N=N_F=20$ particles (+). For the pure TG gas, the comparison to the fitting function $\sim|x|^{-\alpha}$ (double-dotted line) gives $\alpha=0.44$, a value close to the one found for the homogeneous case where $\alpha=0.5$. For the pure Fermi gas, the comparison to the fitting function $\sim \sin(2k_F x)/|x|^{\alpha}$ (dot-dashed line) gives $\alpha=0.93$, a value close to the one found for the homogeneous case where $\alpha=1$. Panel (d) is the same as panel (c) but in linear scales.}
 \end{figure}

In Fig.~\ref{fig0}, we display the off-diagonal behaviour of 
%the fermionic one-body density matrix, rescaled by the square root of the density profiles, i.e. we plot the fermionic one-body correlation function
$g_B^{(1)}(x,y)$ and $g_F^{(1)}(x,y)$ as obtained for $y=-x$. 
Fig.~\ref{fig0}a shows the results for a Bose-Fermi mixture with $N=20$ particles when the boson concentration is varied while Fig.~\ref{fig0}c and Fig.~\ref{fig0}d show the same quantity for a pure TG gas and a pure Fermi gas. At large distances, the fermionic one-body correlator of the BF mixture displays a power-law decay $|x|^{-\alpha}$ 
modulated by typical fermionic oscillations.
The wave vector of the oscillations is close to (but not exactly equal to) $2 k_F=2\pi n_F(0)$ as evidenced in Fig.~\ref{fig0}d.
The power-law exponent $\alpha$ depends on the boson concentration in the mixture as is illustrated in Fig.~\ref{fig0}b. For $N_B=N_F=N/2$ the exponent is $\alpha\simeq 2.6$, i.e. larger than that of a pure Fermi gas, where $\alpha=1$ (for a homogeneous gas in the thermodynamic limit), as well as that of a pure TG gas, where   $\alpha=1/2$ (again for a homogeneous gas in the thermodynamic limit). We therefore find that the Bose-Fermi interactions strongly affect the one-body correlations at large distance.
\section{Density profiles}
\label{sec:density}
\begin{figure}
 \includegraphics[width=0.7\linewidth,clip=true]{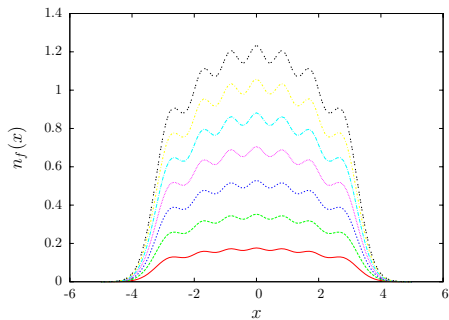}
 \caption{\label{fig1}(Color online) Fermionic density profiles $n_F(x)$ (in
units of $1/\ell_0$) versus $x$ (in units of $\ell_0$), as
 obtained for a BF mixture with a fixed total number of $N=7$ particles when the number of fermions is increased from $N_F=1$ (solid line) up to $N_F=7$ (double-dashed line). Normalizing each curve by $N_F/N$ would collapse them on a same and single curve.}
 \end{figure}

Since the square of an anti-symmetric operator is the unity,
the bosonic and fermionic density profiles 
$n_B(x) = \rho_B^{(1)} (x,x)$ and $n_F(x) = \rho_F^{(1)} (x,x)$ are exactly the same up to normalization factors. Since $\rho_B^{(1)}$ is proportional to $\rho_{TG}$, we have that $n_B(x)$ and $n_F(x)$ are both proportional to the spatial density $n_{TG}(x) = \rho_{TG}(x,x)$ of a harmonically trapped TG gas made of $N=N_B+N_F$ bosons \cite{Girardeau2007}:
\begin{eqnarray}
\frac{n_B(x)}{N_B} = \frac{n_F(x)}{N_F} = \frac{n_{TG}(x)}{N}.
%n_{B,F}(x;N_{B,F}) & = & \frac{N_{B,F}}{N}\, n_{TG} (x;N)\, .
\end{eqnarray}
The main consequence of this result is that strongly interacting
spin-polarized fermions and impenetrable bosons 
do not display any phase separation (or spatial demixing).
Both the fermionic and the bosonic density profiles present a number
of peaks equal to the total number of atoms and the position of the peaks
is exactly the same for bosons and fermions. As an illustrative example, we plot in Fig.\ \ref{fig1} the fermionic density profiles obtained for a BF mixture with a fixed total number of $N=7$ particles when the number of fermions is increased from $N_F=$1 up to $7$.

We also note in passing that for a purely fermionic ensemble, the
Hamilton differential operator does not distinguish positions and momenta up to scaling. Hence, for noninteracting fermions the functional dependence of the momentum distribution \textit{vs} momentum $p$ is exactly the same as the functional dependence of the density profile \textit{vs} $x$ (dashed curve in Fig.\ \ref{fig1} for the case of 7 fermions).  In the next section we discuss how this functional dependence is affected by the interaction with the bosons.

\section{Momentum Distribution}
\label{sec:momentum}
Using the scaled variables, the momentum distribution is defined as the Fourier transform of the one-body
density matrix according to:
\begin{equation}
n(p) =  \frac{1}{2 \pi} \int_{-\infty}^\infty dx  \int_{-\infty}^\infty dy\, \rho^{(1)}(x,y) \, e^{-ip(x-y)}.
\label{def_mom}
\end{equation}

\begin{figure}
\includegraphics[width=0.7\linewidth]{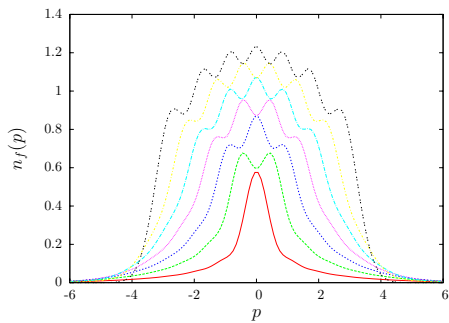}%
\caption{\label{fig2} (Color online) Fermionic momentum distribution $n_F(p)$
(in units of $1/p_0$) versus $p$ (in
units of $p_0$), as obtained for a BF mixture with a fixed total number of $N=7$ particles when the number of fermions is raised from $N_F=$1 (solid line) up to $N_F=$7 (double-dashed line). As fermions are added, $N_F$ "fermionic" oscillations develop in the momentum distribution.}
\end{figure}

The case for bosons is easily cleared. Indeed, since $\rho_B^{(1)}$ and $\rho_{TG}$ are proportional, $n_B(p)$ is identical to $n_{TG}(p)$ (up to normalization). But the case for fermions is more involved because the expression \eqref{rhofermion} for the density matrix could not be reduced to a simpler analytical formula, and we have resorted to numerical computations.  Fig.\ \ref{fig2} shows our results for a BF mixture with a fixed total number of $N=7$ particles when the number of bosons is increased from $N_B=0$ up to 6. Fermionic oscillations are suppressed and  the tails of the distribution become more prominent. The large-$p$ behaviour of the distributions will be analyzed here below. 

\begin{figure}
\includegraphics[width=0.7\linewidth]{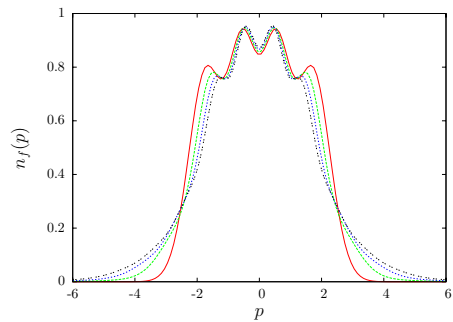}%
\caption{\label{fig3}(Color online) 
Fermionic momentum distribution $n_F(p)$ (in units of $1/p_0$) versus $p$ (in
units of $p_0$), at fixed number of fermions $N_F=4$  when bosons are gradually added. The curves range from $N_B=0$
(solid line) down to $N_B=3$ (double-dashed line). As one can see,
adding more and more bosons smoothens the fermionic oscillations.}
\end{figure}

To evidence how the presence of bosons can influence the spread of $n_F(p)$, we have also considered an ensemble of $N_F=4$ fermions and we have gradually added bosons to it. In Fig.\ \ref{fig3} we have plotted the corresponding fermionic momentum distributions when $N_B$ is increased from $0$ up to $3$. Adding more and more bosons smoothens the fermionic oscillations while the tails of the distribution get enhanced and the distribution broadens. 
This effect can be understood by looking at the asymptotic behaviour of
the fermionic momentum distribution  when $p \to \infty$.
%At fixed number of fermions, the spread of the fermionic momentum distribution increases with the number of bosons.

%\subsection{Asymptotic behaviour}

To this purpose, we use the alternative expression
\begin{equation}
n_F(p) = N_F \int {\rm d}Y \, |\hat\Psi_0 (Y,p)|^2
\end{equation}
where $Y$ is a shorthand for the collection of all $x_i$ variables except the last fermionic one $x_N$ (i.e. $i = 1, \cdots N-1$). Here $\hat\Psi_0 (Y,p)$ stands for the Fourier transform, evaluated at $p$, of the many-body wavefunction $\Psi_0(Y, x_N)$ with respect to its last fermionic variable $x_N$
\begin{equation}
\hat\Psi_0 (Y,p) = \int_{-\infty}^\infty \frac{dx_N}{\sqrt{2\pi}} \, \Psi_0 (Y,x_N) \, e^{-ipx_N}.
\end{equation}
We now use the mathematical result that for any \( f(z) = |z-z_0|^{\alpha-1}\,
F(z)\), where $F(z)$ is a regular function and \( \alpha >0 \), \( \alpha
\neq 1,3,5,\cdots \), we have 
\begin{equation}\label{eqn:result}
\lim_{|k|\to \infty}\!\!\left[|k|^\alpha \!\!\int_{-L/2}^{L/2}\!\!dz\, e^{-ik(z-z_0)} f(z) \right]\!=\! 2 F(z_0)\cos\left(\frac{\pi\alpha}{2}\right)
\, \Gamma(\alpha).
\end{equation}
Following the procedure outlined in Ref. \cite{Olshanii2003}, we next obtain the large-$p$ asymptotics of $\hat\Psi_0 (Y,p) $
\begin{equation}
\hat\Psi_0 (Y,p) \sim \dfrac{-2c_N}{\sqrt{2\pi}\, p^2} \, f_0'(Y) \, K(Y) \, 
e^{-\sum_{j=1}^{N-1} x_j^2/2}
\label{psitrans}
\end{equation}
where $f_0'$ is given by \eqref{thiseq} but with the fermionic index now only running in $\mathcal{F}' = \mathcal{F}\!\setminus\!\!\{N\}$ and
\begin{equation}
K(Y) = \sum_{j\in \mathcal{B}} e^{-i px_j - x_j^2/2} \prod_{\begin{array} {c} \scriptstyle l \in \mathcal{B} \\[-1ex] \scriptstyle l \not=j \end{array}} \prod_{n \in \mathcal{F}'} \, |x_j-x_l| (x_n-x_j).
\end{equation}

The momentum distribution then follows by squaring and integrating over $Y$. After some manipulations, we get $n_F(p) \sim C p^{-4}$ where the constant $C$ is given by
\begin{equation}
\label{C1}
C = \frac{2N_F N_B}{\pi} \, c_N^2 \, \int_{-\infty}^\infty dx \, e^{-2x^2} D(x).
\end{equation}
The function $D(x)$ reads
\begin{equation}
D(x) = \!\! \int \!\! dU \prod_{i=3}^N \Big[ e^{-x_i^2}\prod_{j=3}^{i-1} (x-x_i)^4(x_i-x_j)^2\Big],
\end{equation}
$U$ being the shorthand for the collection of all $x_i$ for $3 \leq i \leq N$. This expression can be compacted to a determinant form by noting that $D(x) = (N-2)! \det G(x)$ where $G(x)$ is the square matrix of order $(N-2)$ with entries
\begin{equation}
G_{jk}(x) = \int_{-\infty}^\infty\! dt\,  e^{-t^2}(x-t)^4 \, t^{j+k-2}.
\end{equation}
Each of these matrix elements can be evaluated using Gamma functions 
\cite{Forrester2003}. As a net result the coefficient $C$ can alternatively be written as
\begin{equation}
\label{C2}
C = \frac{2N_F N_B}{\pi} \, (N-2)! \, c_N^2 \, \int_{-\infty}^\infty dx \, e^{-2x^2} \det G(x).
\end{equation}

\begin{figure}[htb]
\includegraphics[width=0.8\linewidth]{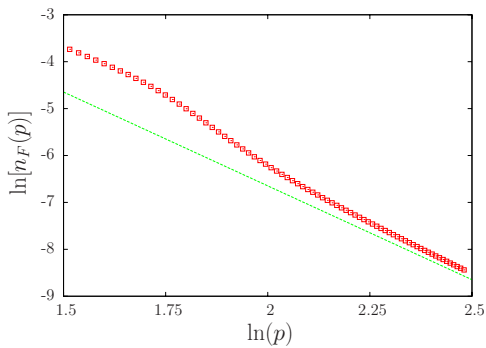}%
\caption{\label{fig6} (Color online) Momentum distribution $n_F(p)$ (in units of $1/p_0$) as a function of $p$ (in units of $p_0$) for the case of a BF mixture made of $N_F=2$ fermions and $N_B=5$ bosons. The plot shows the comparison between the analytical asymptotic prediction $Cp^{-4}$, with $C$ being given by \eqref{C2}, (dashed line) and the exact result computed from Eq. (\ref{def_mom}) (squares).}
\end{figure}

The asymptotic result $n(p)\sim C p^{-4}$ captures well the large-p behaviour of $n_F(p)$. This is illustrated in a log-log graph in Fig.\ \ref{fig6}, where a good agreement is displayed between the asymptotic result and the full calculation of the momentum distribution (\ref{def_mom}).

The $p^{-4}$ asymptotic behaviour of the fermionic momentum distribution 
in the BF mixture under consideration is the same as that of a 
TG gas \cite{Minguzzi2002}. 
It is straightforward to show that \eqref{C1} or \eqref{C2} can indeed be rewritten as
\begin{equation}
C = \lim_{p\rightarrow\infty}p^4n_F(p)=\dfrac{N_F\,N_B}{N(N-1)}\lim_{p\rightarrow\infty}p^4n_{TG}(p)
\label{code}
\end{equation}
where $n_{TG}(p)$ is the momentum distribution for a harmonically trapped
TG gas of $N$ identical bosons. The factor $N_F/N$ ensures that $n_F(p)$ is indeed normalized to the total number $N_F$ of fermions in the system, while the remaining factor $N_B/(N-1)$ gives the weight
of the bosonic contribution at large $p$. We can also notice that, for a BF mixture with a fixed number $N$ of particles, the effect of the bosons on the fermionic distribution will be maximized when the number of bosons equals the number of fermions, i.e. $N_F=N_B$.

Interestingly enough, the coefficient $C$ can be related to the two-body density matrix $\rho^{(2)}$. The two-body density matrix is the same for both the bosonic and fermionic components as it does not depend on the sign of the many-body wavefunction, and coincides with the two-body density matrix of an ideal Fermi gas in the same external potential. It is defined as 
\begin{equation}
\rho^{(2)}(x,y;x',y') \!\!= \!\!N(N-1) \!\!\!\int \!{\rm d}Z \Psi_0 (Z,x,y)\Psi_0^* (Z,x',y'),
\end{equation}
where $Z$ is a shorthand for the collection of all $x_j$ except the last two fermionic variables (i.e. $j= 1, \cdots, N-2$).  Using the explicit definition of $\rho^{(2)}$ for the mixture we can rewrite \eqref{C1} as 
\begin{equation}
C=\frac{2N_FN_B}{\pi N(N-1)}
 \int_{-\infty}^\infty\! {\rm d}x
\lim_{x',x''\rightarrow x} \frac{\rho^{(2)}(x',x;x'',x)}{|x-x'|
| x-x''|}.
\label{mom_ro2}
\end{equation}
Using \eqref{code}, we also obtain the large-momentum behaviour of a pure TG gas as  
\begin{equation}
\label{eq:asTG}
\lim_{p\rightarrow\infty}p^4n_{TG}(p) = 
\frac{2}{\pi} \int_{-\infty}^{+\infty}\!\!\! {\rm d}x\,\lim_{x',x''\rightarrow x} \frac{\rho^{(2)}(x',x;x'',x)}{|x-x'|
|x-x''|}.
\end{equation}
 This expression is the TG limit extension of the results found by Olshanii \cite{Olshanii2003}
in the case of a homogeneous gas of hard-core bosons on a wire of length $L$ 
with \textit{finite} interactions
described by the 1D $s$-wave scattering length $a_{1D}$, namely
\begin{equation}
n_p = \frac{4}{(p/\hbar)^4}\frac{\rho^{(2)}(0,0;0,0)}{a_{1D}^2}.
\end{equation}
Here $p$ is given in full units and not rescaled to the harmonic  
oscillator
ones. Furthermore, the distribution of discrete momenta
$n_p$ is normalized as $\sum_{j=-\infty}^{\infty} n_{2 \pi \hbar j/L}=N$.
 In the TG limit, both the 1D scattering length  and the diagonal element of the two-body correlation function 
$\rho^{(2)}(0,0;0,0)$ vanish; our expression (\ref{eq:asTG}) gives the corresponding limiting value.

\section{Expansion}
\label{sec:exp}
We now turn to the dynamical evolution of the harmonically trapped BF mixture when the trap is switched off by turning down to zero the trap frequency according to $\omega(t) = \omega_0 f(t)$ with some known  function $f(t)$. In a real 3D experiment, a 1D expansion could be generated by turning off only the longitudinal confinement and the mixture would then expand in a 1D geometry. This dynamics can be
described exactly by using time-dependent
coherent states \cite{Popov1970} in close analogy with the dynamics
of a TG gas \cite{Minguzzi2005}.
To this purpose we introduce a scaling transformation, acting on both the spatial 
and temporal coordinates, which provides an exact solution describing this expansion in terms of the time-dependent orbitals $u_j(x,t)$ with energy $E_j$
\begin{equation}
u_j(x,t) =  \!\frac{1}{\sqrt{b(t)}}\, 
u_j\!\left(\frac{x}{b(t)}, 0\right) \exp \!\!\left[ i
  \frac{\dot{b}x^2}{2b}\! - \!i E_j \tau(t) \right].
\label{t-orbitals}
\end{equation}
The scaling factor $b(t)$ obeys the ordinary differential equation
$\ddot{b}+ f^2 b= b^{-3}$ with initial conditions
$b(0)=1$ and $\dot{b}(0) = 0$. Finally the rescaled time $\tau$ is given by $\tau(t)= \int_0^t \!dt'[b(t')]^{-2}$.

By exploiting the time-dependent Bose-Fermi mapping \cite{Girardeau2000}, we construct
the many-body wave-function in terms of the orbitals $u_j(x,t)$, and hence we 
obtain the time evolution of the one-body density matrix in the following scaling form \cite{Minguzzi2005}
\begin{equation}
\rho^{(1)}(x,y;t) = \frac{\rho^{(1)}(x,y;0)}{b(t)}\, \exp\left[- \frac{i\dot{b}(x^2- y^2)}{2b(t)}\right].
\end{equation}
The phase acquired by the one-body density matrix is crucial for the time evolution of the momentum distribution, which is determined by
\begin{equation}
n(p,t) = b \!\!\int \!\!{\rm d}x\, {\rm d}y\, \rho^{(1)}(x,y;0) e^{-ib\dot{b}(x^2-y^2)/2 - bp(x-y)}.
\end{equation}

\begin{figure}
\includegraphics[width=0.9\linewidth]{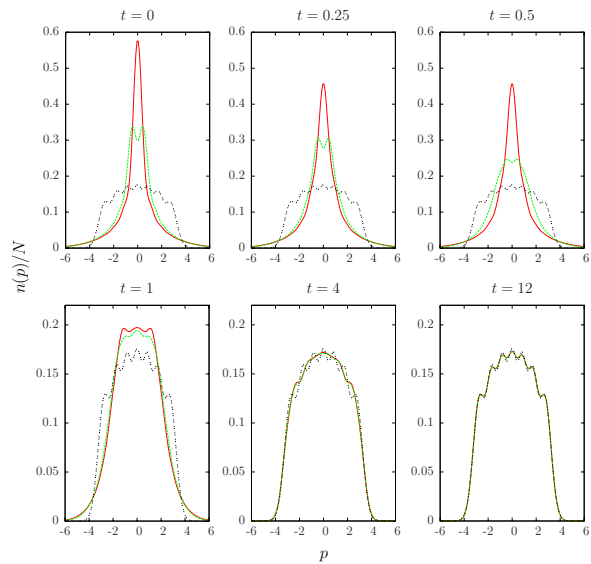}%%%
\caption{\label{fig4} (Color online) Scaled momentum distribution $n(p)$ 
(in units of $1/p_0$) versus $p$ (in units of $p_0$) of an expanding mixture at 
different times (in units of $\omega_0^{-1}$) as indicated in the panels. The different lines correspond to
$N_B=7$ bosons (continuous red lines), $N_F=2$ fermions mixed with $N_B=5$ bosons (dashed green lines), and $N_F=7$ fermions (double-dashed black lines). The distributions have been rescaled by the total number  $N$ of particles in order to be normalized to unity.}
\end{figure}

We now analyze the case of a sudden turn-off of the trap, i.e. $f(t) = 1-\Theta(t)$, $\Theta$ being the step function. The scaling parameter $b(t)$ is then found to be $b(t)= \sqrt{1 + t^2}$. In Fig.\ \ref{fig4} we compare the time evolution of the momentum
distributions obtained for $N_B=7$ bosons in the absence of fermions, for $N_F=2$ fermions 
mixed with $N_B=5$ bosons and for $N_F=7$ fermions in the absence of bosons. 
%renormalized to two particles.
The agreement among all three momentum distributions at large times (bottom-right frame) provides clear evidence for the dynamical fermionization at work in the BF mixture, hence generalizing the results for a pure TG gas obtained in \cite{Rigol2005,Minguzzi2005}.

\section{Summary and concluding remarks}
\label{sec:concl}
We have determined exactly the groundstate properties of
a strongly interacting mixture composed of non-interacting fermions
and hard-core point bosons 
with mutual repulsive point-like hard-core interactions, subjected to a
1D harmonic confinement. Using a generalized Bose-Fermi mapping \cite{Girardeau2007} designed to extend the Bethe Ansatz approach \cite{Imambekov2006} to inhomogeneous confinements, we have explicitly determined the bosonic and fermionic one-body density
matrices. The bosonic one-body density matrix is proportional
to that of a pure Tonks-Girardeau gas whereas the fermionic one can be
given a form suitable for practical calculations by expressing it in terms
of a one-dimensional integral involving the determinant of known special functions.
Knowing the one-body density matrices, we have studied
in detail the equilibrium density profiles, the momentum distributions and
the behaviour of the latter under one-dimensional expansion. Concerning the density profiles, at
variance with the mean-field predictions, no phase separation occurs in the strongly
interacting regime. Both the fermionic and bosonic density profiles are proportional 
to each other and display a shell structure with oscillations 
where the number of peaks is equal to the total number of particles in the mixture. 
While the bosonic momentum distribution keeps proportional to that of
a pure TG gas, we have found that the fermionic momentum distribution shows instead features
of a Fermi sphere at small momenta and slowly decaying tails, like $Cp^{-4}$, at large
momenta. This is a characteristic feature due to the interactions of the fermions with their bosonic partners. We have also determined analytically the coefficient $C$ in terms of the two-body density matrix. This coefficient might be measured in an expansion experiment as suggested  by Tan \cite{Tan2005} in the 3D case.
We have also studied the behaviour of the mixture after a sudden turn-off of the harmonic trap. We have found that the momentum
distribution of the mixture "fermionizes" during the expansion: its long-time limit shape is that of a non-interacting Fermi gas with a {\em total} number $N=N_B+N_F$ of particles. Our predictions could be verified by the on-going experiments on ultracold atomic mixtures loaded in optical lattices in the limit of low filling factor.

%We intend to look into the situation where the contact interaction is finite,
%and both the mass and trap frequency for bosons and fermions are different.  
\acknowledgments
The authors thank B.-G. Englert for
his interest and support in the work.
AM would like to thank Marvin Girardeau for useful discussions and BF would like to thank B. Gr\' emaud for his help with the numerics.
BF and ChM acknowledge support from a PHC Merlion grant (SpinCold 2.02.07) and from CNRS PICS 4159. AM acknowledges funding through the MIDAS-STREP project. This work is supported by the 
National Research Foundation \& Ministry of Education, Singapore.

\end{document}